\documentclass[%
reprint,
superscriptaddress,
altaffilletter,
showpacs,
showkeys,
amsmath,
amssymb,
aps,
prl,
floatfix,
]{revtex4-1}

\usepackage{graphicx}% Include figure files
\usepackage{dcolumn}% Align table columns on decimal point
\usepackage{bm}% bold math
\usepackage{comment}% commenting blocks
\usepackage[usenames,dvipsnames]{color}
\usepackage[symbol]{footmisc}
\usepackage{hyperref}% add hypertext capabilities
\hypersetup{colorlinks = true, allcolors = blue}
\usepackage[mathlines]{lineno}% Enable numbering of text and %displaymath
\usepackage{multirow}
\DeclareGraphicsExtensions{.pdf,.png,.jpg}

\newcommand{\SeNucleai}{3.88~$\times~10^{25}$}

\newcommand{\exposure}{9.95 kg~$\times$~yr}

\newcommand{\startDAQ}{June 2017} 
\newcommand{\stopDAQ}{December 2018} 
\newcommand{\se}{$^{82}$Se}

\newcommand{\onu}{$0\nu\beta \beta$}
\newcommand{\tnu}{$2 \mathrm{\nu\beta \beta}$}

\newcommand{\acpt}{$\mathring{a}_{\text{of}}^{(3)}$}
\newcommand{\adir}{$\bm{a}_{\text{of}}^{(3)}$}
\newcommand{\cupid}{\ensuremath{\rm CUPID\hbox{-}0}}

\begin{document}

\title{First search for Lorentz violation in double beta decay with scintillating calorimeters}

% All university affiliations addresses go here:
\newcommand{\sapienza}{\affiliation{Dipartimento di Fisica, Sapienza Universit\`a di Roma, P.le Aldo Moro 2, 00185, Rome, Italy}}
\newcommand{\infnroma}{\affiliation{INFN Sezione di Roma, P.le Aldo Moro 2, 00185, Rome, Italy}}
\newcommand{\lnl}{\affiliation{INFN  Laboratori Nazionali di Legnaro, I-35020 Legnaro (Pd) - Italy}}
\newcommand{\lngs}{\affiliation{INFN  Laboratori Nazionali del Gran Sasso, I-67010 Assergi (AQ) - Italy}}
\newcommand{\lbl}{\affiliation{Lawrence Berkeley National Laboratory , Berkeley, California 94720, USA}}
\newcommand{\infnge}{\affiliation{INFN  Sezione di Genova, I-16146 Genova - Italy}}
\newcommand{\unige}{\affiliation{Dipartimento di Fisica, Universit\`{a} di Genova, I-16146 Genova - Italy}}
\newcommand{\infnmib}{\affiliation{INFN  Sezione di Milano - Bicocca, I-20126 Milano - Italy}}
\newcommand{\unimib}{\affiliation{Dipartimento di Fisica, Universit\`{a} di Milano - Bicocca, I-20126 Milano - Italy}}
\newcommand{\csnsm}{\affiliation{CSNSM, Univ. Paris-Sud, CNRS/IN2P3, Universit\'e Paris-Saclay, 91405 Orsay, France}}
\newcommand{\cea}{\affiliation{IRFU, CEA, Universit\'e Paris-Saclay, F-91191 Gif-sur-Yvette, France}}
\newcommand{\gssi}{\affiliation{Gran Sasso Science Institute, 67100, L'Aquila - Italy}}
\newcommand{\usc}{\affiliation{Department of Physics  and Astronomy, University of South Carolina, Columbia, SC 29208 - USA\label{USC}}}
\newcommand{\fier}{\affiliation{Finnish Institute for Educational Research, P.O.Box 35 FI-40014 University of Jyv\"askyl\"a, Finland}}
\newcommand{\ctp}{\affiliation{Center for Theoretical Physics, Sloane Physics Laboratory, Yale University, New Haven, Connecticut 06520-8120, USA}}

\author{O.~Azzolini}\lnl
\author{J.W.~Beeman}\lbl
\author{F.~Bellini}\sapienza\infnroma
\author{M.~Beretta\email[Corresponding author: ]{mattia.beretta@mib.infn.it}}\unimib\infnmib
\author{M.~Biassoni}\infnmib
\author{C.~Brofferio}\unimib\infnmib
\author{C.~Bucci} \lngs
\author{S.~Capelli}\unimib\infnmib
\author{L.~Cardani}\infnroma
\author{P.~Carniti}\unimib\infnmib
\author{N.~Casali}\infnroma
\author{D.~Chiesa}\unimib\infnmib
\author{M.~Clemenza}\unimib\infnmib
\author{O.~Cremonesi}\unimib\infnmib
\author{A.~Cruciani}\infnroma
\author{I.~Dafinei}\infnroma
\author{S.~Di~Domizio}\unige\infnge
\author{F.~Ferroni}\gssi\infnroma
\author{L.~Gironi}\unimib\infnmib
\author{A.~Giuliani}\csnsm
\author{P.~Gorla} \lngs
\author{C.~Gotti}\unimib\infnmib
\author{G.~Keppel}\lnl
\author{M.~Martinez}\thanks{Present Address: Fundaci\'on ARAID and Laboratorio de F\'isica Nuclear y Astropart\'iculas, Universidad de Zaragoza, C/ Pedro Cerbuna 12, 50009 Zaragoza, Spain}\sapienza\infnroma
\author{S.~Nagorny}\altaffiliation[Present Address: Queen's University, Physics Department, K7L 3N6, Kingston (ON), Canada]{}\gssi\lngs
\author{M.~Nastasi}\unimib\infnmib
\author{S.~Nisi}\lngs
\author{C.~Nones}\cea
\author{I.~Nutini}\unimib\infnmib
\author{D.~Orlandi}\lngs
\author{L.~Pagnanini}\unimib\infnmib
\author{M.~Pallavicini}\unige\infnge
\author{L.~Pattavina} \lngs
\author{M.~Pavan}\unimib\infnmib
\author{G.~Pessina}\infnmib
\author{V.~Pettinacci}\infnroma
\author{S.~Pirro}\lngs
\author{S.~Pozzi}\unimib\infnmib
\author{E.~Previtali}\unimib\infnmib
\author{A.~Puiu}\unimib\infnmib
\author{C.~Rusconi}\lngs\usc
\author{K.~Sch\"affner}\gssi\lngs
\author{C.~Tomei}\infnroma
\author{M.~Vignati}\infnroma
\author{A.~Zolotarova}\altaffiliation[Present Address: CSNSM, Univ. Paris-Sud, CNRS/IN2P3, Universit\'e Paris-Saclay, 91405 Orsay, France]{}\cea

\date{\today}

\begin{abstract}
We present the search for Lorentz violation in the double beta decay of \se~with CUPID-0, using an exposure of \exposure.
We found no evidence for the searched signal and set a limit on the isotropic components of the Lorentz violating coefficient of $\mathring{a}_{\text{of}}^{(3)} < 4.1\cdot10^{-6}$ GeV (90\% Credible Interval).
This results is obtained with a Bayesian analysis of the experimental data and fully includes the systematic uncertainties of the model. This is the first limit on \acpt~obtained with a scintillating bolometer, showing the potentiality of this technique.
\end{abstract}

\maketitle

%%%%%%%%%%%%%%%%%%%%%%%%%%%%%%%%%%%%%%%%%%%%%%%%%%%%%%%%%%%%%%%%
The development of a coherent theory capable of unifying quantum mechanics and general relativity is a central goal of contemporary particle physics. Different solutions to this problem hypothesize the existence of unconventional physical phenomena at the Planck scale ($\sim10^{19}$~GeV), violating the fundamental symmetries of nature. In particular, several models include the breakdown of Lorentz and CPT (Charge-Parity-Time reversal) symmetries for the sake of a consistent quantum gravity description \cite{QuantumSpacetimePhenomenology}. Since this new phenomenology arises at unreachable energies, a direct observation cannot be performed. 
Nevertheless, such new physics can impact the Standard Model (SM) predictions as
an effective theory characterised by Lorentz symmetry violation (LV) \cite{RelViolAndBD}, producing sizeable effects in low energy processes like double beta decay. The Standard Model Extension (SME) \cite{CPTVAndSM,SME_CPTV,GravityAndSM} is the framework where these effective quantum field operators are described, including both LV and CPT-odd operators. LV is included with background fields with non-zero vacuum expectation values, resulting in the sponaneous breaking of space-time symmetry \cite{LorentzViolation}. Neutrino physics is an ideal benchmark to test SME prediction, as many operators affect macroscopic phenomena such as neutrino oscillations \cite{SME_Table}. In particular there exists a \emph{countershaded} operator with no impact on neutrino velocities, that cannot be investigated in neutrino oscillations or time-of-flight measurements. The \emph{countershaded} operator has mass dimension 3, is renormalizable and is odd under CPT. The strenght of its interaction is given by the four independent components of $(a_{\text{of}}^{(3)})^{\alpha}$ coefficient. The absolute value of these components is proportional to the intensity of LV. The 3 directional components are labelled as \adir, while the anisotropic component as \acpt. The former can be studied in experiments directly sensitive to the particle directions, while the latter when directions are not taken into account. The interactions of neutrinos with this operator modify their quadrimomentum from $q^{\alpha} = (\omega,\bm{q})$ to $\Tilde{q}^{\alpha} = ( \omega,\bm{q} + \bm{a}_{\text{of}}^{(3)}- \mathring{a}_{\text{of}}^{(3)}\bm{\hat{q}} )$ \cite{CPTfromBD,LorentzViolation}. In double beta decay (\tnu) experiments measuring only the energy of the two emitted electrons, only \acpt~remains as possible source for LV.The standard \tnu~electrons sum spectrum \cite{2nu_HalfLives,2nu_PhaseSpace,2nu_Matrix} is modified in shape, with a sizable modification parameterized by \acpt.
In this work we present the search for this deformation in CUPID-0, exploiting the excellent resolution and background rejection capability of our detector to put a limit on the value of \acpt.

%%%%%%%%%%%%%%%%%%%%%%%%%%%%%%%%%%%%%%%%%%%%%%%%%%%%%%%%%%
CUPID-0 is an experiment designed to search for the neutrinoless double beta decay (\onu) of \se~($Q_{\text{value}}=2997.9\pm0.3$~keV \cite{2nuQVAL}) with a calorimetric approach, using the technique of scintillating bolometers. The detector, described in detail in Ref. \cite{CUPIDDETECTOR}, is composed by 26 ZnSe scintillating crystals (24 enriched at 95\% in \se~and two with natural isotopic abundance) acting as bolometers and interleaved with high purity Germanium bolometric light detectors (LD).
The experiment is operating at a base temperature of 10~mK in the hall A of Gran Sasso National Laboratory (Italy). The ZnSe crystals and LD are held in position and thermalised through a
mechanical copper structure and PTFE supports. The crystals are surrounded by Vikuity$^{\text{TM}}$ reflective foil, to enhance the light collection.
The dual heat/light readout allows us to combine the excellent energy resolution of bolometers with the background rejection capabilities of scintillators.
Indeed, CUPID-0 reached the lowest background ever measured in a \onu~bolometric detector, setting the most stringent limits in the search of the \se~\onu~to the fundamental and excited states of $^{82}$Kr \cite{CUPIDPRL,CUPIDPRL2,EXCITED_STATES}.
A comprehensive background model has also been developed for CUPID-0 \cite{BKGmodel}, evaluating and localizing all the possible sources of background for the detector.
The understanding of the experimental data obtained with this model allowed to exploit the high number of \tnu~events for detailed studies on this process.
CUPID-0 is therefore a suitable candidate with which to perform the study of \tnu~spectral shape to evaluate the \acpt~parameter. 

%%%%%%%%%%%%%%%%%%%%%%%%%%%%%%%%%%%%%%%%%%%%%%%%%%%%%%%%%%
\begin{figure}[htp]
\centering
\includegraphics[width=\columnwidth]{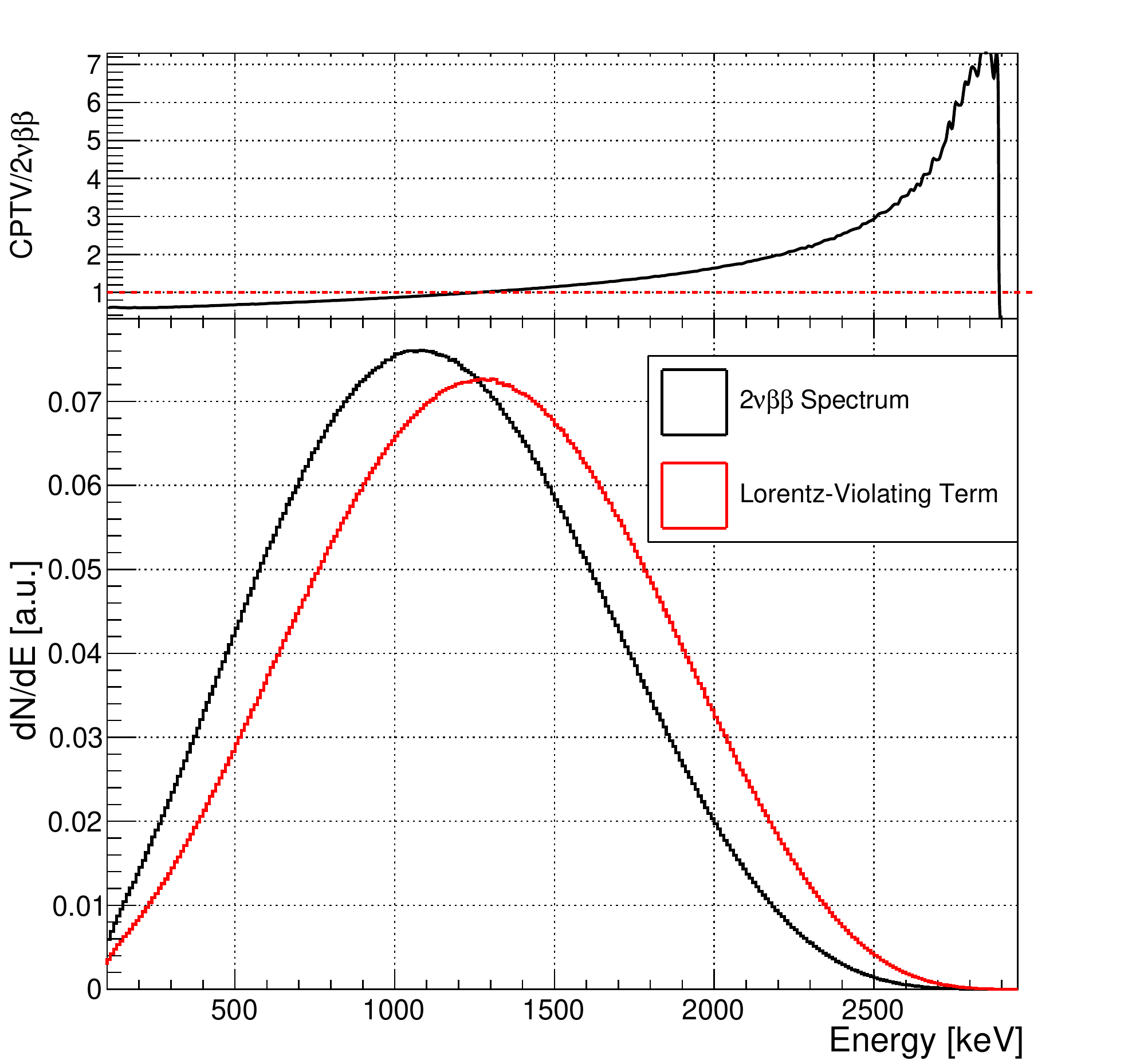}
\caption{Comparison between standard (black) and Lorentz violating (red) \tnu~simulated electron spectra for \se. The simulation is based on exact phase space calculation for \tnu{} \cite{2nu_PhaseSpace}. The emitted electrons are propagated in the detector geometry and the bremsstrahlung emission is also implemented. The spectra are normalized to the same integral. In the upper panel the ratio of the two spectra is reported as a function of the energy.} 
\label{fig.2NuComp}
\end{figure}

The inclusion of Lorentz violating \tnu~changes the energy spectrum of the two emitted electrons with an additive term parameterized by \acpt. The differential spectrum can be expressed with

\begin{align}
\nonumber \frac{d\Gamma}{dE} = &C\cdot F(Z,t_{1})\sqrt{t_{1}(t_{1}+2)}(t_{1}+1) \cdot\\
 \nonumber & F(Z,E-t_{1})\sqrt{E-t_{1}(E-t_{1}+2)}(E-t_{1}+1) \cdot \\
 \nonumber &[(Q-E)^5+10\cdot\mathring{a}_{\text{of}}^{(3)}(Q-E)^4)] \\
 =&C\cdot\bigg(\frac{dI^{\text{Theo}}_{2\nu,\text{SM}}}{dE}+10\cdot \mathring{a}_{\text{of}}^{(3)}\frac{dI^{\text{Theo}}_{2\nu,\text{LV}}}{dE}\bigg)
\label{eq.2NuSpectrum}
\end{align}

where $t_{1}$ is the energy of one of the two emitted electrons, $E$ is the sum of the two emitted electrons kinetic energy, $Q$ is the Q-value of the \tnu, $C$ is the factor taking into account the nuclear matrix element and normalization constants \cite{CPTfromBD,CPTV-EXO}, $F$ is the Coulomb correction \cite{2nu_Factor} and $\frac{dI^{\text{Theo}}_{2\nu,\text{SM}}}{dE}$ and $\frac{dI^{\text{Theo}}_{2\nu,\text{LV}}}{dE}$ are the SM and LV terms of the decay amplitude. The LV is represented as an additive term, characterized by a different spectral shape and whose weight is given by \acpt.
In figure \ref{fig.2NuComp} the simulations of the two \tnu~modes for \se~are reported. These simulations take into account all the inefficiencies in the two electron collection, such as the detector geometry and the bremmstralhung emission. The \acpt~parameter can be quantified comparing the respective integral of the two spectra. 
The integration of Eq. \ref{eq.2NuSpectrum} gives the prediction of the respective weight of the two decay modes in terms of \acpt:
\begin{linenomath*}
\begin{align}
    &C\cdot\int_{0}^{Q}dE{\frac{dI^{\text{Theo}}_{2\nu,\text{SM}}}{dE}} = C\cdot I^{\text{Theo}}_{2\nu,\text{SM}} \label{eq.TheoWeight1}\\
    &C\cdot10\cdot \mathring{a}_{\text{of}}^{(3)}\cdot\int_{0}^{Q}dE{\frac{dI^{\text{Theo}}_{2\nu,\text{LV}}}{dE}} =C\cdot10\cdot \mathring{a}_{\text{of}}^{(3)}I^{\text{Theo}}_{2\nu,\text{LV}}
\label{eq.TheoWeight2}
\end{align}
\end{linenomath*}
where $I^{\text{Theo}}_{2\nu,\text{SM}}$ and $I^{\text{Theo}}_{2\nu,\text{LV}}$ are the prediction for the standard and Lorentz violating \tnu~modes respectively.
The left side of Eq.s \ref{eq.TheoWeight1},\ref{eq.TheoWeight2} can be measured experimentally, and the ratio between the two relations provides a pathway for the evaluation of \acpt. Defining $\Gamma^{\text{Exp}}_{2\nu,\text{SM}}$ and $\Gamma^{\text{Exp}}_{2\nu,\text{LV}}$ respectively as the standard and Lorentz violating \tnu~measured decay rates, \acpt~can be calculated from:

\begin{equation}
   \mathring{a}_{\text{of}}^{(3)}
   = 
   \frac{1}{10}\frac{I^{\text{Theo}}_{2\nu,\text{SM}}}{I^{\text{Theo}}_{2\nu,\text{LV}}}
   \cdot
   \frac{\Gamma^{\text{Exp}}_{2\nu,\text{LV}}}{\Gamma^{\text{Exp}}_{2\nu,\text{SM}}} 
   \label{eq.acptCalc}
\end{equation}

In this work, the predicted values for $I^{\text{Theo}}_{2\nu,\text{SM}}$ and $I^{\text{Theo}}_{2\nu,\text{LV}}$ have been calculated from the integration of Eq.s\ref{eq.TheoWeight1},\ref{eq.TheoWeight2}. The ratio is independent from the matrix element used, since the Lorentz violation in SME arises from a momentum modification. No particular addition had therefore to be taken into account to adapt this evaluation to \se.  

On the measurement side, the evaluation of the standard and Lorentz violating \tnu~has been performed with a Bayesian procedure on CUPID-0 experimental data, with a fit of simulated background spectra to the measured data. Data used in this work have been collected from \startDAQ~to \stopDAQ~and correspond to a total Zn$^{82}$Se exposure of \exposure~(\SeNucleai~$^{82}$Se nuclei~$\times$~yr). 

Time, amplitude, pulse shape, light output and other pulse information were extracted from the collected events following the procedure described in \cite{ANALISI_CUPID}. Selection cuts were applied to the data to exclude non-particle events, with a total efficiency of $\epsilon=(95.7\pm0.5)\%$, constant above 150~keV \cite{BKGmodel}.
The searched signal consists of two electrons detected by a single crystal. Other event categories can be used to constrain the background sources, such as multi-site events used to constraint muo-indiced showers \cite{BKGmodel}.
Events were then classified according to the nature of the interacting particle ($\alpha$ or $\beta/\gamma$) and the number of ZnSe crystals that simultaneously triggered the event.
The particle identification was exploited only above 2~MeV, where the measurement of the different light output allowed to distinguish $\alpha$ from $\beta/\gamma$ particles \cite{ANALISI_CUPID,BKGmodel}. Below 2~MeV, the poor resolution of light detectors prevented such discrimination. The residual $\alpha$ particles that could not be identified below 2~MeV were added to the $\beta$/$\gamma$ spectrum, both in measured and simulated data \cite{BKGmodel}. 
Counting the number of crystals fired simultaneously in a coincidence window of 20~ms allowed to define events in which one, two or multiple crystals were involved simultaneously. Due to its small statistics, the last category was only used to constrain the background produced by muons. The particle identification and the event multiplicity were used to construct four spectra: single hit spectra of $\alpha$ and $\beta/\gamma$ particles ($\mathcal{M}_{1\alpha}$ and $\mathcal{M}_{1\beta/\gamma}$), double hit spectrum ($\mathcal{M}_{2}$) and the double hit spectrum where the energy is the sum of the two hit energy ($\Sigma_{2}$). The same spectra have also been defined on simulated data, allowing the fit procedure. As previously explained, the large majority of the \tnu~signal events occurs in $\mathcal{M}_{1\beta/\gamma}$.
To describe these spectra, CUPID-0 background model uses 33 different sources, identified on the basis of experimental data and previous experimental results obtained in the same infrastructure (CUORE-0 experiment \cite{CUORE0BKGMod}).
For the analysis described in this paper, the Lorenz violating \tnu~was added to the background model as an additional contribution. The background model was then constructed by a simultaneous fit of the four experimental spectra with a linear combination of the Monte Carlo spectra obtained for the 33+1 sources. 
The free parameters of the fit are the activities of each source, parameterized as the coefficients of this linear combination.
The measured spectrum of each source is simulated by the means of a Geant4 based Monte Carlo simulation, taking into account the detector geometry and its finite temporal and energetic resolution \cite{BKGmodel}. The simulation software also addresses the bremsstrahlung emissions of the electrons, which affects the \tnu~spectral shape.
A Bayesian approach is chosen to solve this problem, \cite{BKGmodel,CUORE0BKGMod} hence a prior distribution for each normalization parameter is defined. The priors of the 33 background components are the same of \cite{BKGmodel}. For the Lorentz violating \tnu~a non-negative uniform prior has been chosen, considering this process as an alternative decay channel with respect to the standard \tnu.
The joint posterior probability density function (pdf) of all fit parameters is sampled with JAGS (Just Another Gibbs Sampler) \cite{JAGSman}, a software based on a Markov Chain Monte Carlo algorithm. The joint posterior pdf is subsequently marginalized to obtain the pdf for each normalization parameter. 
This strategy exploits the relevant experimental signatures of the different background sources to constraint their activities. Both \tnu~modes produce most events in the $\mathcal{M}_{1\beta/\gamma}$. As a direct consequence, the background sources which are constrained by other spectra, or whose normalization is anchored to a peak in the experimental spectrum, are not affected by the introduction of Lorentz violating \tnu~in the model. 
%quali sorgenti non sono toccate dal 2nu
The unaffected sources are:
\begin{itemize}
    \item both bulk and surface $\alpha$ sources localized in the ZnSe crystals and in the reflective foil, since their normalization is constrained by the $\mathcal{M}_{1\alpha}$ spectrum;
    \item $\gamma$ sources whose normalization is determined by the intensity of the experimental peaks;
    \item muons, since they are normalized on the higher multiplicity spectra.
\end{itemize}

The remaining contribution to the background model, excluding the two \tnu~modes, are represented by 10 sources correlated to the searched signature, since they produce a continuum in the 1500-2000~keV energy range. Their effect on the measured coupling constant is subsequently discussed.

\begin{figure}[htp]
\centering
\includegraphics[width=\columnwidth]{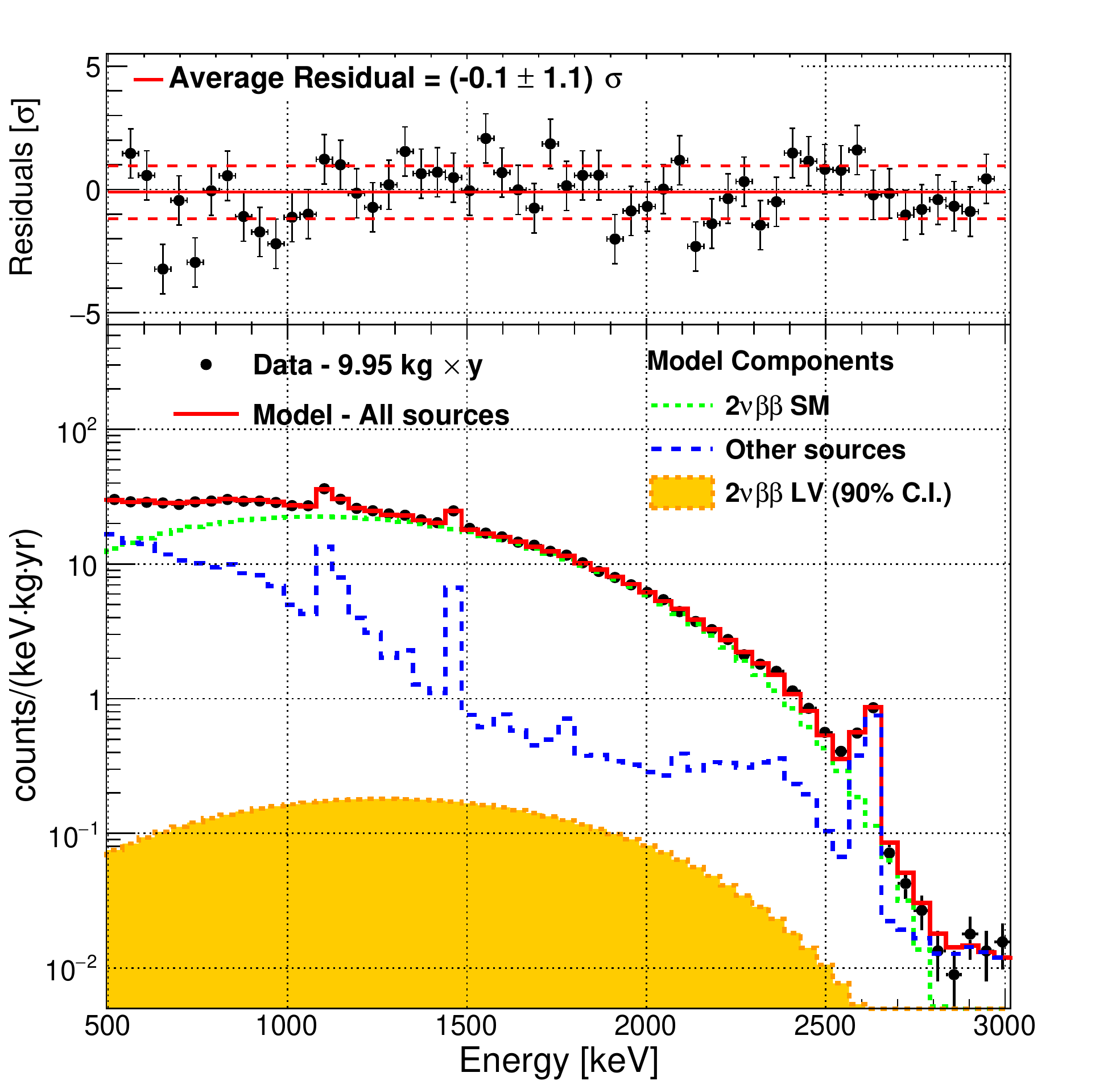}
\caption{
CUPID-0 $\mathcal{M}_{1\beta/\gamma}$ (\exposure~of ZnSe exposure) (black dots). The red line shows the reconstruction performed with the fit. The green dashed line is the reconstruction of the standard \tnu, while the orange spectrum is the 90\% CI limit for the Lorentz violating component.The blue dashed line is the sum of all the other 32 background contributions. In the upper panel, the residuals are reported as a function of the energy. The agreement is good over all energies, expecially around the peaks, where the model exploits the relevant spectral features to fix its reconstruction.
} 
\label{fig.ModelComp}
\end{figure}

To extract the value of \acpt, the ${\Gamma^{\text{Exp}}_{2\nu,\text{LV}}}/{\Gamma^{\text{Exp}}_{2\nu,\text{SM}}}$  ratio
(see Eq.s \ref{eq.TheoWeight1},\ref{eq.TheoWeight2},\ref{eq.acptCalc}) is computed for each sampling of the joint posterior pdf.
In Fig. \ref{fig.ModelComp} the experimental $\mathcal{M}_{1\beta/\gamma}$ spectrum is shown together with the reconstruction. The experimental data are well described by the chosen model, even with the inclusion of the Lorentz violating \tnu. The distribution of fit residuals has a Gaussian shape, with average compatible with zero and $\sigma$ compatible with 1. In addition, the values of the background sources activities present only small variations with respect to the results reported in Ref. \cite{BKGmodel}. In particular, the \tnu~activity obtained by this fit is (9.8$\pm$0.1)$\cdot10^{-4}$~Bq/kg, while in the background model is (9.96$\pm$0.03)$\cdot10^{-4}$~Bq/kg.

%%%%%%%%%%%%%%%%%%%%%%%%%%%%%%%%%%%%%%%%%%%%%%%%%%%%%%%%%%
From the posterior distribution for $\Gamma^{2\nu}_{\text{Standard}}$/$\Gamma^{2\nu}_{\text{LV}}$ ($\mathcal{R}$) the distribution for \acpt~can be calculated, combining the sampled posterior distribution with the theoretical value for the reciprocal weight of the two decay modes (Eq.s \ref{eq.TheoWeight1}, \ref{eq.TheoWeight2}, \ref{eq.acptCalc}). The conversion factor from $\mathcal{R}$ to \acpt~is calculated from Eq.s~\ref{eq.2NuSpectrum}, \ref{eq.TheoWeight1}, \ref{eq.TheoWeight2} as:
\begin{equation}
  \frac{I^{\text{Theo}}_{2\nu,\text{SM}}}{10 \cdot I^{\text{Theo}}_{2\nu,\text{LV}}}
  =
  (213.3\pm0.7)\cdot10^{-6}~\text{GeV}
   \label{eq.TheoRatioVal}
\end{equation}
where the error is due to the uncertainty on the Q-value of \se~\tnu. We folded this uncertainty in $\mathcal{R}$, using the Gaussian distribution for $I^{\text{Theo}}_{2\nu,\text{SM}}/(10 \cdot I^{\text{Theo}}_{2\nu,\text{LV}})$. 
Since no significant evidence of the Lorentz violating \tnu{} could be detected, a 90\% credible interval (CI) limit is determined from the obtained \acpt{} distribution.
\\
The posterior pdf is affected by the correlation with the nuisance parameters of the model, i.e. the other normalization coefficients and the influence variables. 
The correlation with other model parameters is taken into account during the marginalization of the joint posterior distribution.
The influence variables instead are arbitrary parameters used in the fit and have to be changed to determine their effect on the analysis result. The bin width used to build the spectra and the lower threshold applied to the data have been considered as influence variables. The following tests have been performed:
\begin{itemize}
    \item \textbf{Bin:} bin values of 15~keV, 30~keV and 50~keV have been used to perform the fit;
    \item \textbf{Threshold:} thresholds of 200~keV, 300~keV and 500~keV have been used in different fits.
\end{itemize}

Alongside the influence variables, the hypothesis on the source location in the background model constitutes another possible source of systematic uncertainty. In particular, the positioning of $^{40}$K, $^{60}$Co and $^{232}$Th/$^{238}$U has to be taken into account, since these sources produce experimental signatures correlated to the Lorentz violating \tnu.
As reported in Ref. \cite{BKGmodel}, the CUPID-0 cryostat model is radially divided by the Roman lead shield in two sections, one internal and one external. $^{40}$K, $^{60}$Co and $^{232}$Th/$^{238}$U can be present both inside and outside the Roman lead shield. 
The $^{232}$Th/$^{238}$U component can in addition be localized in the Roman lead shield, providing further variability. During the performed tests, each source has been removed from one of the possible locations, resulting in two tests (internal or external) for $^{40}$K and $^{60}$Co and three tests for $^{232}$Th/$^{238}$U (no internal, no external, no Roman lead).
An additional influence on the limit also comes from the presence of an unidentified contamination of pure $\beta$-emitters. In particular, from the $\beta$-decaying isotopes with negligible $\gamma$ emission, long half life (>100~d) and high $Q_{\text{value}}$. The only isotope simultaneously meeting these requirements is the $^{90}$Sr, a fission product originating the $\beta$ decay sequence $^{90}$Sr $\rightarrow$ $^{90}$Y $\rightarrow$ $^{90}$Zr, with $Q_{\text{value}}$ of 546~keV and 2281~keV respectively. To evaluate the effects of this possible contamination, a test has been performed including $^{90}$Sr in the list of sources.
As reported in \cite{CUPIDPRL2}, the energy calibration is affected by bias evaluated over all the interest energies. To control the effects of this bias on the current analysis, an evaluation of the obtained limit is performed using the corrected energy scale.
For each test, the model shows a satisfactory agreement with data. The fit residual distribution can always be modelled with a gaussian with mean value compatible with 0 and $\sigma$ compatible with 1.

\begin{figure}[htp]
\centering
\includegraphics[width=\columnwidth]{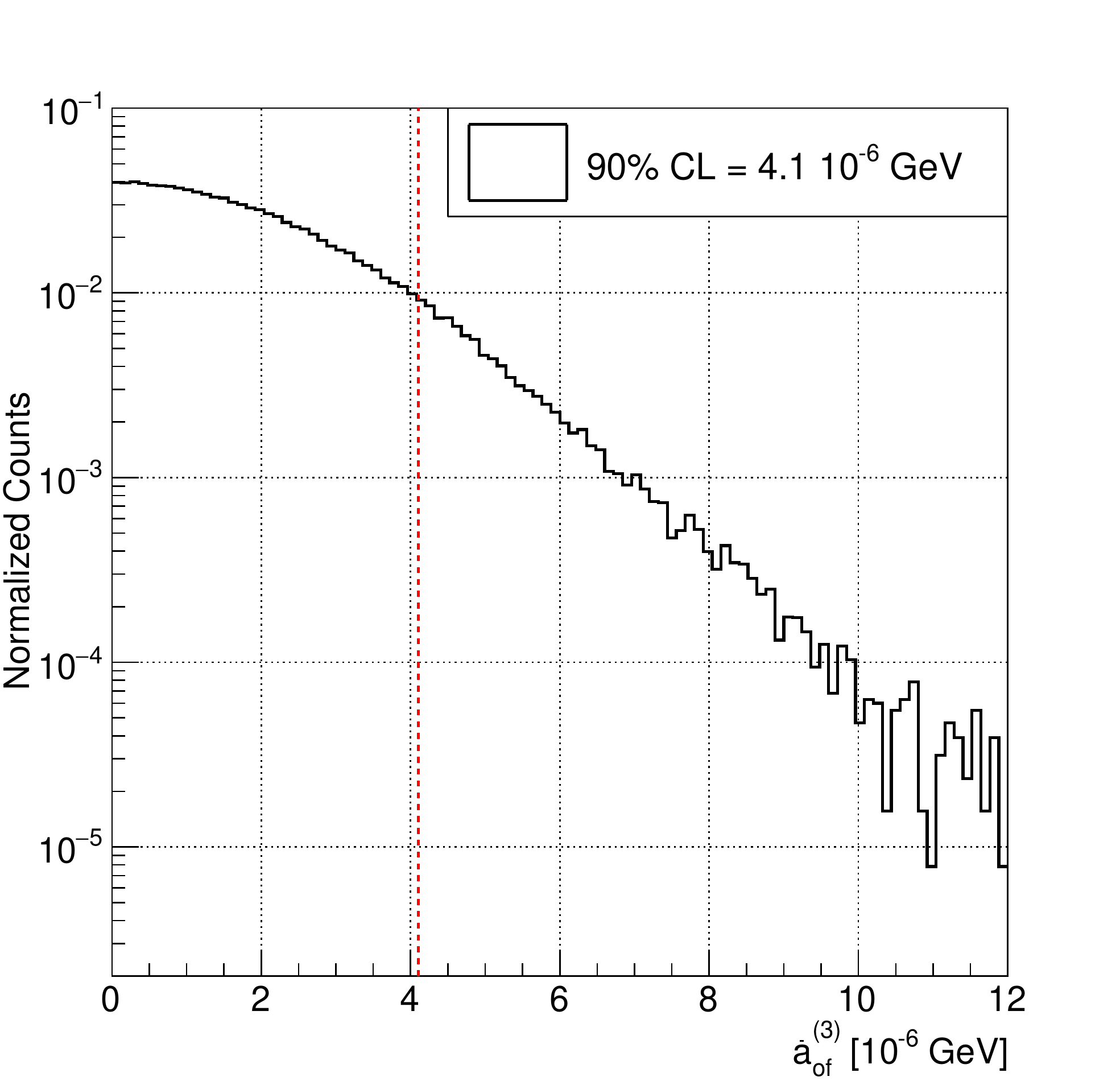}
\caption{Total posterior probability distribution for \acpt. The dashed line represents the 90\% credible interval, corresponding to $\mathring{a}_{\text{of}}^{(3)} < 4.1\cdot10^{-6}$ GeV.} 
\label{fig.A0Limit}
\end{figure}

The results of the different tests have been combined in each category by adding the relative posterior distribution functions for \acpt, according to the law of total probability. A uniform prior for each test has been considered, resulting in an average of the distributions in each test family. The corresponding 90\% CI limits on \acpt~are reported in Tab. \ref{table:SystTest}. 
To obtain a final limit taking into account all the studied effects, the posteriors for each test family have been added with equal weight. The final limit has the value of $\mathring{a}_{\text{of}}^{(3)} < 4.1\cdot10^{-6}$~GeV. The final posterior distribution is shown in figure \ref{fig.A0Limit}, with the evaluated 90\% CI limit.

%table fo results
\begin{table}[htb]
\begin{center}
\caption{Results of different tests performed to evaluate the systematics effects on the \acpt~limit. For each row, different values have been tested and combined adding the relative posterior distribution. The total result is obtained as a limit on the sum of all the family posteriors with equal weight.}
\label{table:SystTest}
\begin{tabular}{cc}
\hline
Variable & Result [10$^{-6}$ GeV]\\
\noalign{\smallskip}\hline\noalign{\smallskip}
\noalign{\smallskip}
\multicolumn{2}{c}{Influence variables}\\
\noalign{\smallskip}
binning & $<$3.7 \\
\noalign{\smallskip}
threshold & $<$3.5 \\
\noalign{\smallskip}
\noalign{\smallskip}\hline\noalign{\smallskip}
\multicolumn{2}{c}{Different Models}\\
\noalign{\smallskip}
$^{40}$K & $<$4.0 \\
\noalign{\smallskip}
$^{60}$Co & $<$3.6 \\
\noalign{\smallskip}
$^{232}$Th and $^{238}$U & $<$4.2 \\
\noalign{\smallskip}
$^{90}$Sr & $<$5.8 \\
\noalign{\smallskip}
\noalign{\smallskip}\hline\noalign{\smallskip}
\multicolumn{2}{c}{Energy scale uncertainty}\\
\noalign{\smallskip}
Calibration & $<$4.1 \\
\hline
Total & $<$4.1 \\
\hline
\end{tabular}\\
\end{center}
\end{table}

The obtained result establishes a bound for \acpt, obtained for the first time with the shape analysis of the \tnu~spectrum measured with a scintillating bolometer. 
Even with a limited exposure of \exposure, the performances of CUPID-0 scintillating bolometers allowed to reach a limit competitive with previously published ones by EXO-200 collaboration ($\mathring{a}_{\text{of}}^{(3)} < 7.6 \cdot 10^{-6}~\text{GeV} $, with an exposure of 100 kg$\cdot$yr) \cite{CPTV-EXO}
and NEMO-3 collaboration ($\mathring{a}_{\text{of}}^{(3)} < 3.5 \cdot 10^{-7}~\text{GeV}$, with 34.4 kg$\cdot$yr) \cite{CPT_NEMO-3}.
The potentiality of Bayesian analysis applied to bolometric experimental data, showed in \cite{BKGmodel,CUORE0BKGMod}, is further established.
\\
This result proves that scintillating bolometers can perform spectral shape studies with high sensitivity, even when using a limited statistics. As a consequence, the development of high exposure detectors based on this technique can provide tools to overcome current detectability limits. In addition, this approach can be used to study the Lorentz violation in different \tnu~decaying isotopes. Changing the studied crystals, in fact, allows to perform the same analysis on different isotopes, with comparable sensitivities \cite{NutiniPhD}.  
\\
\begin{acknowledgments}
This work was partially supported by the European Research Council (FP7/2007-2013) under contract LUCIFER no. 247115. The work of JK was supported by the Academy of Finland (Grant No. 314733). We thank M. Iannone for his help in all the stages of the detector assembly, A. Pelosi for constructing the assembly line, M. Guetti for the assistance in the cryogenic operations, R. Gaigher for the mechanics of the calibration system, M. Lindozzi for the cryostat monitoring system, M. Perego for his invaluable help in many tasks, the mechanical workshop of LNGS (E. Tatananni, A. Rotilio, A. Corsi, and B. Romualdi) for the continuous help in the overall set-up design. We acknowledge the Dark Side Collaboration for the use of the low-radon clean room. This work makes use of the DIANA data analysis and APOLLO data acquisition software which has been developed by the CUORICINO, CUORE, LUCIFER and, \cupid~collaborations. This work makes use of the Arby software for Geant4 based Monte Carlo simulations, that has been developed in the framework of the Milano – Bicocca R\&D activities and that is maintained by O. Cremonesi and S. Pozzi.
\end{acknowledgments}
\bibliography{biblio}
\end{document}